\documentclass[letterpaper,twocolumn,10pt]{article}
\usepackage{usenix-2020-09}

\usepackage{tikz}
\usepackage{amsmath}
\usepackage{filecontents}
\usepackage{xspace}
\usepackage[normalem]{ulem}
\usepackage{graphicx}
\usepackage{subcaption}
\usepackage[most]{tcolorbox}
\usepackage{booktabs}
\usepackage{enumitem}
\usepackage{amssymb}

\newtcolorbox{takeaway}{
  colback=gray!15,    
  colframe=gray!50,   
  boxrule=0.5pt,      
  arc=3pt,            
  left=4pt, right=3pt, top=3pt, bottom=1pt 
}

\definecolor{darkred}{RGB}{180,0,0}
\definecolor{darkblue}{RGB}{0,0,160}
\newcommand{\red}[1]{\textcolor{darkred}{#1}}
\newcommand{\blue}[1]{\textcolor{darkblue}{#1}}

\begin{document}

\date{}

\title{\LARGE \bf Fantasy: \Large Efficient Large-scale Vector Search on GPU \\
Clusters with GPUDirect Async}

\author{
{\rm Yi Liu, Chen Qian} \\
 \it University of California Santa Cruz
}

\maketitle

\newcommand{\sys}{\textbf{Fantasy}\xspace}

\begin{abstract}
Vector similarity search has become a critical component in AI-driven applications such as large language models (LLMs). To achieve high recall and low latency, GPUs are utilized to exploit massive parallelism for faster query processing. However, as the number of vectors continues to grow, the graph size quickly exceeds the memory capacity of a single GPU, making it infeasible to store and process the entire index on a single GPU. Recent work uses CPU-GPU architectures to keep vectors in CPU memory or SSDs, but the loading step stalls GPU computation.
We present \textbf{\sys}, an efficient system that pipelines vector search and data transfer in a GPU cluster with GPUDirect Async. \sys overlaps computation and network communication to significantly improve search throughput for large graphs and deliver large query batch sizes.
\end{abstract}

\section{Introduction}

Vector similarity search has become a foundational technology in modern AI systems~\cite{distributedann,dhnsw,milvus,hnsw,zhao2020song,diskann}, enabling efficient similarity retrieval among high-dimensional embeddings. It plays a central role in retrieval-augmented generation (RAG) for large language models~\cite{ragcache,ragdoll,rago}, recommendation systems~\cite{yang2025gpu}, multimodal understanding~\cite{ERNIE-4.5}, and semantic search. Vector similarity search is particularly important because it determines how quickly and accurately a system can identify the most relevant data points from massive embedding collections, directly affecting the quality and responsiveness of AI applications. As these applications scale to billions of embeddings, approximate nearest neighbor search (ANNS) has emerged as a fundamental research problem~\cite{jang2023cxl,zhang2023fast,zhang2024fast,guo2025achieving,hnsw,dhnsw,singlestore}, aiming to efficiently retrieve the top-k most closest vectors with high recall and low latency under limited computation and memory resources.

Among various ANNS approaches, graph-based indexing structures have demonstrated decent performance in balancing recall and latency. These structures organize vectors as nodes in a navigable graph, where edges connect nearby vectors based on similarity. For example, Hierarchical Navigable Small World (HNSW) builds multi-layer proximity graphs to enable efficient search~\cite{hnsw,dhnsw}. During query processing, HNSW performs a greedy style traversal, starting from an entry node and iteratively moving toward more similar vectors. This search strategy allows the algorithm to quickly reach the top-k nearest neighbors, achieving high recall with low query latency.

To improve search parallelism, recent studies have explored graph-based vector search for GPU acceleration, such as CAGRA in cuVS~\cite{cagra}. These approaches exploit GPU to accelerate neighbor exploration and distance computation across large batches of queries. While they significantly reduce query latency, they still suffer from scalability limitations as the graph size grows with the number of vectors. 
For example, indexing datasets like SIFT1B with CAGRA can consume hundreds of gigabytes of memory, far exceeding the capacity of a single GPU’s HBM (typically around 40-80 GB). Even if part of the graph is stored in host memory or SSD and fetched on demand, the data transfer overhead constrains it from supporting a large batch size.

\begin{figure}[!t]
  \centering
  \begin{subfigure}[b]{0.49\textwidth}
    \centering
    \includegraphics[width=\textwidth]{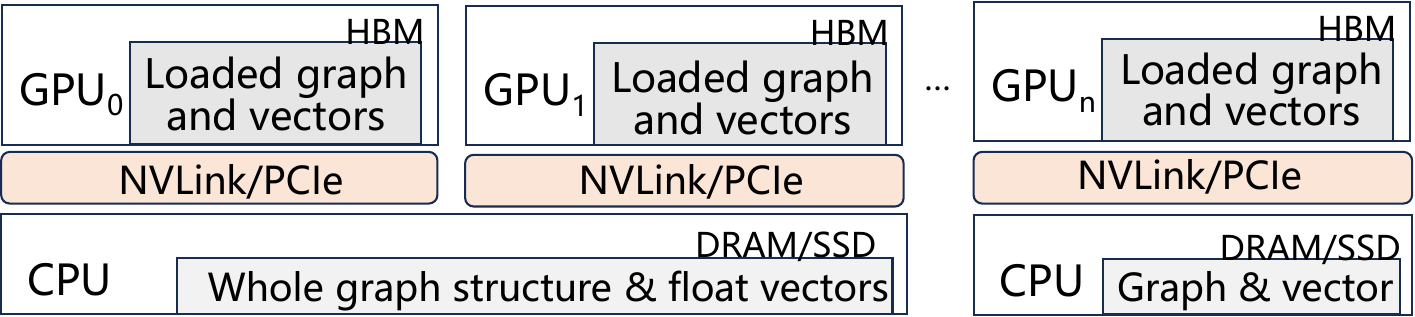}
    \caption{\red{Out-of-Core} search approach.}
    \label{figu:intro:arch0}
  \end{subfigure}
  \begin{subfigure}[b]{0.49\textwidth}
    \centering
    \vspace{1ex}
    \includegraphics[width=\textwidth]{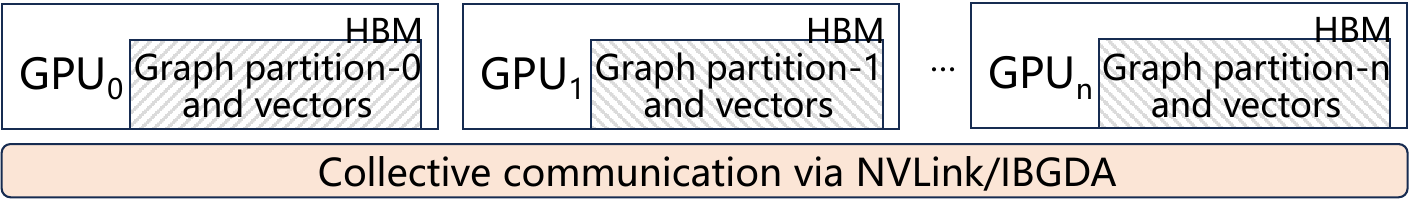}
    \caption{\blue{In-Memory} collective search approach.}
    \label{figu:intro:arch1}
  \end{subfigure}
  \vspace{-3.5ex}
  \caption{GPU-based vector search architectures.}
  \label{fig:intro:arch}
  \vspace{-2.5ex}
\end{figure}

Recently, the InfiniBand GPUDirect Async (IBGDA) transport enables direct and efficient GPU-to-GPU communication across nodes by eliminating the CPU from the data movement control path.~\cite{nvshmem}
IBGDA allows the GPUs to initiate RDMA operations asynchronously through the NIC. The NIC can directly access GPU memory via GPUDirect RDMA over the PCIe interconnect, achieving data transfers at full RDMA bandwidth (e.g., 400~Gb/s) without going through host memory. Also, the GPU’s streaming multiprocessors (SMs) remain idle during communication because the NIC handles the transfer independently, so that it significantly improves end-to-end throughput between GPUs.

We propose a novel architecture named \sys, which stores large-scale vector database in a distributed GPU cluster. Each GPU holds a partition of both the indexing graph and its corresponding vectors in HBM. 
\sys enables each GPU to accept batched query requests and perform collective communication as well as vector search efficiently across the cluster. 
Since vector search is a memory-constrained task on modern GPUs such as Hopper, \sys achieves high per-GPU throughput even under large batch sizes by fully utilizing HBM and interconnect bandwidth. 
In the following sections, we explain (1) \uline{why \sys outperforms CPU DRAM/SSD-based baselines} and demonstrate (2) \uline{how \sys orchestrate \textit{computation} and \textit{communication} in GPU clusters to execute vector search efficiently.}

\section{Motivation}
\label{section:motivation}

Even when a large vector database and its indexing structure cannot fit on a single GPU, they can be partitioned and stored across multiple GPUs in a distributed manner. This motivates our exploration of new system architectures that coordinate storage and computation across GPUs to sustain high throughput and low latency at scale.

Taking the \textbf{CAGRA} algorithm as an example, the main computational cost of a query comes from computing distances between the query vector and its candidate neighbors traversed during graph search. The symbols used are summarized in Table~\ref{tab:cagra-vars}.

\begin{table}[h]
\centering
\caption{Notation for compute and memory analysis in CAGRA}
\label{tab:cagra-vars}
\resizebox{0.48\textwidth}{!}
{\begin{tabular}{ll}
\toprule
\textbf{Symbol} & \textbf{Definition} \\
\midrule
$L$ & Number of visited nodes per query \\ 
$M$ & Average out-degree (neighbors per node) \\
$d$ & Vector dimension \\
$bt_{\text{elem}}$ & Bytes per vector element  \\ 
$F_{\text{elem}}$ & FLOPs per element for distance computation (typically 2–3) \\
$F_q$ & Total FLOPs per query \\
$bt_q$ & Total bytes read from GPU memory per query \\
$\mathrm{AI}$ & Arithmetic intensity  \\ 
\bottomrule
\end{tabular}
}
\end{table}

Each query computes distances to roughly $L \times M$ candidate vectors, each of dimension $d$.  
The total FLOPs are:
\begin{equation}
  F_q \approx F_{\text{elem}} \cdot L \cdot M \cdot d.
\end{equation}

Each candidate vector is fetched once from GPU memory (HBM), so the data movement per query is:
\begin{equation}
  bt_q \approx (bt_{\text{elem}} \cdot d) \cdot (L \cdot M).
\end{equation}

The arithmetic intensity (ratio of computation to memory traffic) is:
\begin{equation}
  \mathrm{AI} = \frac{F_q}{bt_q}
  \approx \frac{F_{\text{elem}} \cdot L M d}{bt_{\text{elem}} \cdot d \cdot L M}
  = \frac{F_{\text{elem}}}{bt_{\text{elem}}}.
\end{equation}

For FP32 ($bt_{\text{elem}}=4$) and $F_{\text{elem}} \approx 2\text{--}3$, we get:
\begin{equation}
  \mathrm{AI} \approx 0.5 \text{--} 0.75~\text{FLOP/byte}.
\end{equation}

This arithmetic intensity is low, meaning the workload is dominated by memory bandwidth rather than SM compute. Even if FP16 is used, $\mathrm{AI}$ only doubles to around $1$–$1.5$~FLOP/byte, still far below the compute-bound threshold ($\geq 10$).  

\begin{takeaway}
\textbf{Observation:} Graph-based greedy search (e.g., CAGRA~\cite{cagra}) is memory-intensive, as each step performs few arithmetic operations per large amount of data fetched from HBM.
\end{takeaway}

Then, we want to compare the two architectures shown in Figure~\ref{fig:intro:arch}, and compare the upper bound of batch sizes they can support. 
We consider three bandwidth (throughput) ceilings for a single GPU:

\begin{itemize}[leftmargin=2em, topsep=1pt, partopsep=0pt, itemsep=1pt, parsep=2pt]
  \item Compute throughput: $B_{\mathrm{FLOP}}$ (peak TFLOP/s)
  \item HBM bandwidth: $B_{\mathrm{HBM}}$ ($\sim$1.5~TB/s)
  \item Host or disk $\rightarrow$ GPU bandwidth: \\$B_{\mathrm{IO}}$ 
        (PCIe~4 $\times$16 $\approx$ 32~GB/s; PCIe~5 $\times$16 $\approx$ 64~GB/s; 
         NVMe single-drive sequential bandwidth $\approx$ 7~GB/s)
\end{itemize}
\vspace{2ex}

For each \textit{batch} of $bs$ queries, the iteration time can be approximated as follows.
\vspace{-2ex}
\paragraph{(1) Out-of-core search approach.}
\begin{equation}
T_{(1)} \approx 
\max \!\left( 
\frac{bs \cdot F_q}{B_{\mathrm{FLOP}}},\;
\frac{bs \cdot bt_q}{B_{\mathrm{HBM}}},\;
\frac{bs \cdot bt_q}{B_{\mathrm{IO}}}
\right)
\end{equation}

\paragraph{(2) In-HBM collective approach.}
\begin{equation}
T_{(2)} \approx 
\max \!\left( 
\frac{bs \cdot F_q}{B_{\mathrm{FLOP}}},\;
\frac{bs \cdot bt_q}{B_{\mathrm{HBM}}}
\right)
\end{equation}


For graph-based search such as HNSW or CAGRA, the arithmetic intensity (AI) is low, thus typically:
\begin{equation}
\frac{bt_q}{B_{\mathrm{HBM}}} \gg \frac{F_q}{B_{\mathrm{FLOP}}}.
\end{equation}
Hence in In-HBM search, performance is dominated by HBM bandwidth rather than by SM compute.  
In Out-of-core search, an additional slower channel $B_{\mathrm{IO}}$ appears, yielding:
\begin{equation}
T_{(1)} \approx 
\max \!\left(
\frac{bs \cdot bt_q}{B_{\mathrm{HBM}}},\;
\frac{bs \cdot bt_q}{B_{\mathrm{IO}}}
\right)
= \frac{bs \cdot bt_q}{B_{\mathrm{IO}}} \quad 
\end{equation}
since $B_{\mathrm{IO}} \ll B_{\mathrm{HBM}}$.

\begin{figure*}[!t]
    \centering
    \begin{minipage}[b]{0.32\textwidth}
        \centering
        \includegraphics[width=\textwidth]{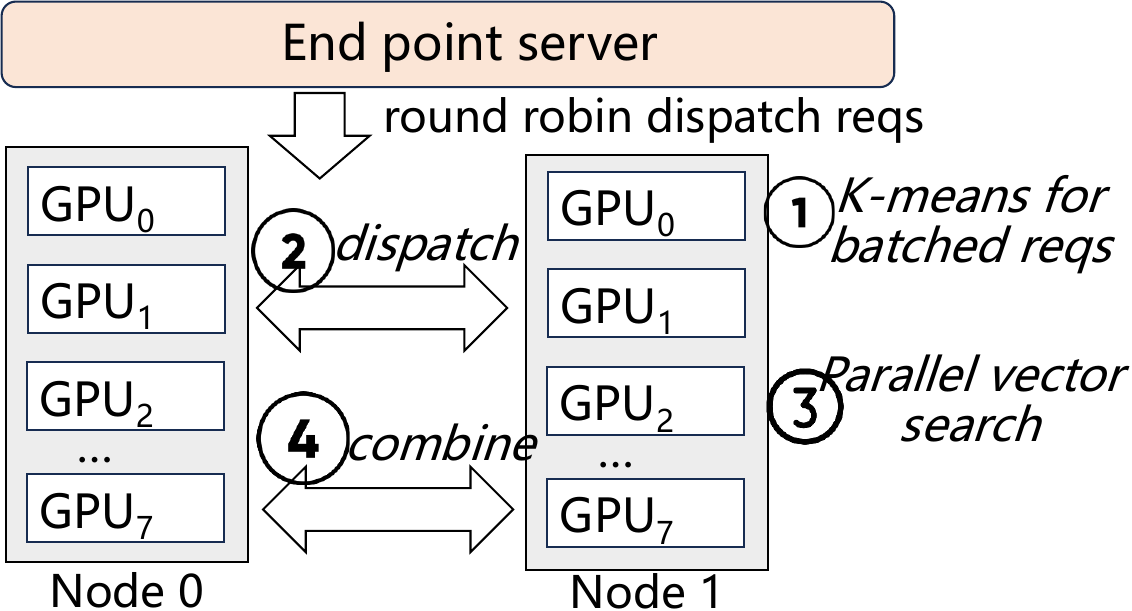}
        \caption{Workflow of vector dispatching and combining in \sys.}
        \label{fig:method:workflow}
    \end{minipage}
    \hspace{-1ex}
    \begin{minipage}[b]{0.675\textwidth}
        \centering
        \includegraphics[width=\textwidth]{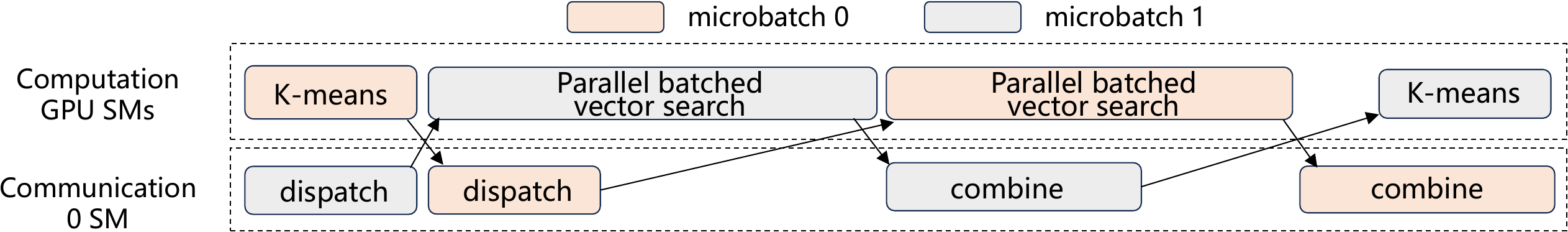}
        \caption{Two microbatches pipeline in \sys.}
        \vspace{5.5ex}
        \label{fig:method:pipeline}
    \end{minipage}
    \vspace{-1.5ex}
\end{figure*}
Thus, with in-HBM collective search, the HBM channel is the first to saturate. Because HBM is fast, a large $bs$ can be sustained before reaching the bandwidth limit. In Out-of-Core search, the PCIe/NVMe channel saturates first; throughput peaks at a much smaller $bs$, and increasing $bs$ further can even reduce performance due to queueing and variance.

\begin{takeaway}
\textbf{Observation:} The In-HBM search architecture can process vector queries with a larger batch size compared to the Out-of-Core approach, thereby achieving higher average search throughput per GPU.
\end{takeaway}

\section{Methodology}
\label{section:methodology}
Our goad of \sys is given there is a queried vector, it will return approximate top-k full float vectors.

\subsection{Vector Search Workflow on GPU Clusters}

To support billion-scale vector search on large graph-based indexes, we partition the global graph into multiple subgraphs and distribute them across GPUs in a cluster. Each rank stores one partition entirely in its HBM, including both the indexing structure and the associated vectors. This design avoids out-of-core data movement during query execution and enables fully in-memory search at high throughput. However, distributing the index requires an efficient mechanism to determine which partitions (or GPUs) should handle each query. We leverage a \textit{K-means}-based clustering scheme to partition the vector space and guide query routing across GPUs.

Figure~\ref{fig:method:workflow} illustrates the overall workflow, which consists of four stages:

\begin{enumerate}[leftmargin=*]
  \item \textbf{K-means as classifier.} Each rank receives a batch of query vectors from the endpoint server in a round-robin fashion. The rank performs \textit{K-means} on GPUs to determine the top-$c$ nearest clusters that each query vector belongs to. These clusters correspond to partitions distributed across the GPU cluster. The result of this stage identifies which ranks are responsible for processing each query vector.

  \item \textbf{Dispatch queried vectors.} After the cluster assignment, each queried vector is dispatched to its corresponding ranks via IBGDA. IBGDA enables direct GPU-to-GPU communication at RDMA bandwidth across nodes without CPU or GPU's SM involvement. This eliminates intermediate host memory copies and achieves near line-rate data transfer efficiency. Each queried vector is typically sent to $c$ clusters for parallel search. Those $c$ clusters can be co-located on one rank or distributed across several ranks.

  \item \textbf{Parallel vector search.} Upon receiving dispatched vectors, each GPU performs a batched vector similarity search in parallel. All indexing structures (e.g., HNSW or CAGRA graphs) and float vectors are fully resident in HBM, ensuring that the entire search process executes without additional data transfers. This stage is memory-bound and highly parallelized across GPU threads, allowing efficient distance computations and top-$k$ selection.

  \item \textbf{Combine vector candidate results.} After local searches complete, each rank returns its top-$k$ matching vectors (not only their IDs, but the float vector) to the originating rank via networks. The original rank then aggregates these partial results to generate the final top-$k$ nearest neighbors.
\end{enumerate}

To maximize GPU-cluster utilization, we consider how to orchestrate the four stages in a two-microbatch pipeline~\cite{deepseek-r1,deepseek-v3} shown in Figure.~\ref{fig:method:pipeline} with an example of Hopper GPU.

\subsection{Two-microbatch Pipelining}

\subsubsection{Stage~1: Local K-means Assignment on Each Rank}

We estimate the local computation time of the K-means assignment stage on each rank for a batch of $bs$ query vectors. 
Each query $\mathbf{q}\in\mathbb{R}^{d}$ is compared against all $C$ cluster centroids to find the top-$c$ nearest clusters, which determine the destination ranks for the subsequent dispatch stage. 

\paragraph{Computation model.}
The Euclidean distance between a query and a centroid can be computed as 
$\|\mathbf{q}-\mathbf{c}\|^2=\|\mathbf{q}\|^2+\|\mathbf{c}\|^2-2\,\mathbf{q}\!\cdot\!\mathbf{c}$.
Then, the dominant operation is a batched matrix multiplication between the query and centroid matrices:
\[
\mathbf{Q}_{b\times d} \times \mathbf{C}_{d\times C} \;\Rightarrow\; \mathbf{D}_{b\times C}.
\]
The total number of FLOPs is approximately:
\[
\text{FLOPs} \approx 2\,b\,d\,C.
\]
Since this stage is compute-bound, its execution time on a single rank can be estimated as:
\[
T_{\text{K-means}} \approx \frac{2\,b\,d\,C}{\eta \cdot P},
\]
where $\eta$ is the effective computational efficiency (typically 0.5--0.7 for large GEMMs), and $P$ is the peak arithmetic throughput of the GPU.


Provided we set the workload as $bs=10k$, $d=1536$, and $C=4096$ and vector value type is TF32, 
the peak throughput of NVIDIA A100 GPU can reach to 156~\text{TFLOP/s}, thus, the total compute cost per rank is:
\[
\text{FLOPs} = 2\times10k\times1536\times4096 \approx 2.52\times10^{11}.
\]
Assuming $\eta=0.6$ for TF32 tensor cores:
\[
T_{\text{K-means}} \approx \frac{1.26\times10^{11}}{0.6\times156\times10^{12}} \approx 1.35~\text{ms}.
\]
Thus, we profile the local K-means assignment per batch on each GPU can be completed in sub-millisecond time.

\subsection{Stage 2: All-to-All Dispatch Latency Estimation}

We consider a distributed vector search setup with \( R = 16 \) ranks across two nodes, 
where each node contains 8 ranks connected via NVLink, and the two nodes are interconnected via RDMA.
Each rank stores $C/R$ partitions of a large database and processes a batch of $bs$ queried vectors.
Each vector is dispatched to its top-\(c\) closest partitions, i.e., top-\(c=3\) ranks, 
which are uniformly and randomly distributed across all ranks.
Even if one of the top-\(c\) destinations resides on the same rank, the data transfer still goes through the NVLink path.

Each rank sends \( bs \cdot c \) vectors in total, and the data volume per rank is:
\[
\text{Data per rank} = bs \cdot c \cdot d \cdot 4 \text{bytes},
\]
where \( d = 1536 \) is the vector dimension and each element is stored in FP32.

For random top-\(c\) selections among 16 ranks (8 per node), the probability that a target rank lies within the same node is:
\[
f_{\text{nv}} \approx \frac{8}{16} = 0.5,
\]
and the inter-node fraction is: $f_{\text{rd}} = 1 - f_{\text{nv}} = 0.5.$

The all-to-all dispatch time per rank can be estimated by:
\[
T_{\text{dispatch}} 
= \frac{bs \cdot d \cdot 4 \cdot f_{\text{nv}}}{BW_{\text{nv}}} 
+ \frac{bs \cdot d \cdot 4 \cdot f_{\text{rd}}}{BW_{\text{rd}}},
\]
where \( BW_{\text{nv}} = 600~\text{GB/s} \) is the NVLink bandwidth
and \( BW_{\text{rd}} = 25~\text{GB/s} \) is the RDMA bandwidth per GPU.

Provided we have \( bs = 10k \), \( c = 3 \), and \( d = 1536 \),
the data volumes per rank are:
\[
\text{Data}_{\text{nv}} = \text{Data}_{\text{rd}} = bs \cdot c \cdot d \cdot 4 \cdot f_{\text{nv}} \approx 91.9~\text{MB},
\]

Hence, the total dispatch time per rank is:
\[
T_{\text{dispatch}} 
\approx \frac{97.3~\text{MB}}{600~\text{GB/s}} 
+ \frac{86.6~\text{MB}}{25~\text{GB/s}}
\approx 3.67~\text{ms}.
\]

while intra-node NVLink transfer contributes negligibly to the total latency.
This estimation assumes ideal bandwidth utilization and no congestion or synchronization overheads.

\subsection{Stage~3: Parallel Vector Search Latency}

For each rank, we consider the time to perform vector search over
$c \times bs$ query vectors using the CAGRA index.
CAGRA is a GPU-optimized graph-based similarity search algorithm that
achieves high throughput by maximizing memory bandwidth utilization.

\vspace{0.3em}

In our analysis, we set 
$BW_{\mathrm{HBM}} = 1.55\times 10^{12}~\text{bytes/s}$ as A100 GPU bandwidth, 
$d_g = 32$ as graph out-degree, $I = 6$ as iterations per search, and beam width per iteration $w$ as 6.

Each query visits approximately:
\[
V = I \times w \times d_g = 6 \times 6 \times 32 = 1152
\]
neighbor vectors, yielding a per-query memory traffic of
\[
\text{Bytes/query} = V \times d \times b = 1152 \times 1536 \times 2  \approx 3.539~\text{MB}.
\]
The query throughput on a single rank is limited by
the available memory bandwidth:
\[
\text{QPS} = \frac{BW_{\mathrm{HBM}}}{\text{Bytes/query}}
\approx 4.37\times10^{5}~\text{queries/s}.
\]

\vspace{0.3em}
\noindent\textbf{Search Time:}
For $c \times bs$ queries per rank:
\[
T_{\text{search}}^{(\text{rank})}
= \frac{c \times bs}{\text{QPS}}
= \frac{c \times bs\times \text{Bytes/query}}{BW_{\mathrm{HBM}}}
\approx 68.5~\text{ms per rank}.
\]

\noindent Hence, for $d=1536$ and FP16 precision on an NVIDIA A100 GPU,
the per-rank CAGRA search time for $c\times bs = 30k$ vectors is approximately
\textbf{68.5~ms}.

\subsection{Stage~4: Combine Candidate Vectors}

In the final stage, each rank gathers the top-$k*c$ candidates for each vector
returned from its top-$c$ expert ranks. This operation resembles an inverse
all-to-all, where every expert rank transmits the query results back to the
originating rank that issued the request.

\[
T_{\text{combine}}
= c \times T_{\text{dispatch}} = 11.01~\text{ms}
\]
when $c$ equals to 3.

\bibliographystyle{acm}
\bibliography{sample}

\end{document}